\documentstyle[preprint,aps]{revtex}
\begin{document}
\draft
\title{MICROSCOPIC EQUATION FOR GROWING INTERFACES IN QUENCHED DISORDERED
MEDIA}
\author{L. A. Braunstein\cite{email}$\dagger$, R. C. Buceta$\dagger$ and
A. D\'{\i}az-S\'anchez$\ddagger$}
\address{$\dagger$ Departamento de F\'{\i}sica, Facultad de Ciencias Exactas y
Naturales, Universidad Nacional de Mar del Plata, Funes 3350,
(7600) Mar del Plata, Argentina\\ $\ddagger$ Departamento de
F\'{\i}sica, Universidad de Murcia, E-30071 Murcia, Espa\~na}

\maketitle

\begin{abstract}
We present the microscopic equation of growing interface with quenched noise
for the Tang and Leschhorn model [L. H. Tang and H. Leschhorn, Phys. Rev. A
{\bf 45}, R8309 (1992)].
The evolution equation for the height, the mean height,
and the roughness are reached in a simple
way. An equation for the interface activity density (or free sites
density) as function of time is obtained.
The microscopic equation allows us to express these equations into
two contributions: the diffusion and the substratum contributions.
All these equations shows the strong interplay between the
diffusion and the substratum contribution in the dynamics.
\end{abstract}
\section{INTRODUCTION}
\label{sec:intro}

The investigation of rough surfaces and interfaces has attracted
much attention, for decades, due to its importance in many fields,
such as the motion of liquids in porous media, growth of bacterial
colonies, crystal growth, etc. Much effort has been done in
understanding the properties in these processes \cite{fam}. When a
fluid wet a porous medium, a nonequilibrium self-affine rough
interface is generated. The interface has been characterized
through scaling of the interfacial width $w=\langle[h_i-\langle
h_i\rangle]^2\rangle^{1/2}$ with time $t$ and lateral size $L$.
The result is the determination of two exponents, $\beta$ and
$\alpha$ called dynamical and roughness exponents respectively.
The interfacial width $w\sim L^\alpha$ for $t\gg L^{\alpha/\beta}$
and $w\sim t^\beta$ for $t\ll L^{\alpha/\beta}$. The crossover
time between this two regimes is of the order of
$L^{\alpha/\beta}$.

The formation of interfaces is determinated by several factors, it is very
difficult to theoretically discriminate all of them.
An understanding of the dynamical nonlinearities, the disorder of the media,
and the theoretical model representing experimental results is difficult
to arrive at due the complex nature of the growth.
The disorder affects the motion of the interface and leads to its roughness.
Two main kinds of disorder have been proposed: the ``annealed'' noise
that depends only of time and the ``quenched'' disorder due to
the inhomogeneity of the media in which the moving phase is propagating.
Some experiments such as the growth of bacterial colonies and the motion
of liquids in porous media, where the disorder is quenched, are
well described by the directed percolation depining model.
This model was proposed simultaneously by Tang and Leschhorn \cite{tang}
and Buldyrev {\sl et al.} \cite{buldyrev}.
Braunstein and Buceta \cite{brauns1} showed that the power law scaling for
the roughness only holds at criticality for $t\ll L$. Also, starting from
the macroscopic equation for the roughness the dynamical exponent has been
theoretically calculated. They found $\beta=0.629$ for the critical value
$q_c=0.539$.

In this paper, we use the TL model in order to investigate
the imbibition of a viscous fluid in a porous media driven by
capillary forces. We write a microscopic equation (ME), starting
from the microscopic rules, for the evolution of the fluid height
as function of time. The ME allows us t0 identify two
contributions that dominates the dynamics of the system, the
``diffusion'' and the ``substratum'' contributions. In this context we study
the mean height speed (MHS), the interface activity density (IAD),
{\sl i.e} the density of actives sites of the interface, and the
roughness as function of time. We show that the diffusion
contribution smooth out the surface for $q$ well below the
criticality but enhances the roughness near the critical value. To
our knowledge, the separation into two contributions for all the
quantities studied in this paper and the important role of
the diffusion contribution to the critical power-law behaviour has
never been studied before.

The paper is organized as follows. In section~\ref{sec:me} we derive
the microscopic equation for the evolution of height for the TL model.
In section~\ref{sec:mh} we separate two contributions of the MHS:
the diffusion and the substratum one. We find a relation between
these contributions that allows us to write an analytical equation for the
IAD. In Section~\ref{sec:ru} the temporal derivative of square interface width as
function of time is derived from the ME and the two contributions are
identified. These two contributions allow us to explain the mechanism of
roughness. Finally, we conclude with a discussion in Section~\ref{sec:con}.

\section{THE MICROSCOPIC MODEL}
\label{sec:me}

In the model introduced by Tang and Leschhorn (TL) \cite{tang} the
interface growth takes place in a square lattice of edge $L$  with
periodic boundary conditions. We assign a random
pinning force $g({\bf r})$ uniformly distributed in the interval
$[0,1]$ to every cell of the square lattice. For a given applied
pressure $p>0$ , we can divide the cells into two groups: those
with $g({\bf r}) \le p$ (free or active cells), and those with
$g({\bf r}) > p$ (blocked or inactive cells). Denoting by $q$ the
density of inactive cells on the lattice, we have $q=1-p$ for $0<
p<1$ and $q=0$ for $p\ge 1$. The interface is specified completely
by a set of integer column heights $h_i$ ($i=1,\dots ,L$). At $t =
0$ all columns are assume to have the same height, equal to zero.
During growth, a column is selected at random, say column $i$, and
compared its height with those of neighbor columns $(i - 1)$ and
$(i + 1)$. The growth event is defined as follow. If $h_i$ is
greater than either $h_{i-1}$ or $h_{i+1}$ by two or more units,
the height of the lower of the two columns $(i-1)$ and $(i+1)$ is
incremented in one (in case of the two being equal, one of the two is 
chosen with equal probability). In the opposite case, $h_i < \min
(h_{i-1},h_{i+1}) +2$, the column $i$ advances by one unit
provided that the cell to be occupied is an active cell. Otherwise
no growth takes place. In this model, the time unit is defined as
one growth  attempt. In numerical simulations at each growth
attempt the time $t$ is increased by $\delta t$, where $\delta t =
1 / L$. Thus, after $L$ growth attempts the time is
increased in one unit. In our simulations we used $L=\;8192$ and a
time interval much less than the crossover time to the static
regime.

We consider the evolution of the height of the $i$-th site for the
process described above. We assume periodic boundary conditions
in a one-dimensional lattice of L sites. At the time $t$ a site $j$ is
chosen at random with probability $1/L$. Let us denote by $h_i(t)$ the
height of the $i$-th generic site at time $t$. The set of $\{h_i,i=1,\dots,L\}$
defines the interface between wet and dry cells. The time evolution for
the interface in a time step $\delta t$ = $1/L$ is
\begin{equation}
h_i(t+\delta t) = h_i(t)+\frac{1}{L} G_i(h_{i-1},h_i,h_{i+1})\;,\label{ev}
\end{equation}
where \begin{equation}
G_i=W_{i+1}+W_{i-1}+F_i(h'_i)\,W_i\;\label{gi},
\end{equation}
with
\begin{eqnarray*}
W_{i\pm 1}&=&\Theta(h_{i\pm 1}-h_i-2)
    \{[1-\Theta(h_i-h_{i\pm 2})]+\delta_{h_i,h_{i\pm 2}}/2\}\;,\nonumber\\
W_{i}&=&1-\Theta(h_i-\min(h_{i-1},h_{i+1})-2)\;.
\end{eqnarray*}
Here $h'_i=h_i+1$ and $\Theta(x)$ is the unit step function defined
as $\Theta(x)=1$ for $x\ge 0$ and equals to $0$ otherwise. $F_i(h_i')$
equals to $1$ if the cell at the height $h_i'$ is active
({\sl i.e.} the growth may occur at the next step) or $0$ if the cell is
inactive. $F_i$ is the interface activity function.
$G_i$ takes into account all the possible ways the site $i$ can grow.
The height in the site $i$ is increased by one with probability
\begin{enumerate}
\item\hspace{.5cm}
$1$ \hspace{1.4cm} if $j=i+1$ and $h_{i+1} \ge h_i + 2$ and $h_i < h_{i+2}$,
\item\hspace{.5cm}
$1/2$ \hspace{1cm} if $j=i+1$ and $h_{i+1} \ge h_i + 2$ and $h_i = h_{i+2}$,
\item\hspace{.5cm}
$1$ \hspace{1.4cm} if $j=i-1$ and $h_{i-1} \ge h_i + 2$ and $h_i < h_{i-2}$,
\item\hspace{.5cm}
$1/2$ \hspace{1cm} if $j=i-1$ and $h_{i-1} \ge h_i + 2$ and $h_i = h_{i-2}$,
\item\hspace{.5cm}
$1$ \hspace{1.4cm} if $j=i$ and $h_i < \min(h_{i-1},h_{i+1}) + 2$ and
$F_i(h_i')=1$.
\end{enumerate}
Otherwise, the height is not increased. The cases (1)-(4) are related
to growth due to the neighbors of the site $i$. We shall call
these mechanisms, growth by ``diffusion''. Notice that these
growths are not related to the disorder of the substratum. The
factor 1/2 takes into account the tie of first-neighbor heights at
the $(i\pm 1)$-th site in the cases (2) and (4). The case (5) is related
to local growth, {\sl i.e.}, if the site $i$ is chosen and the
difference of heights between the $i$-th and the lowest of his
neighbors is less than two, then the height of the chosen site
increases by one provided that the cell above the interface is
active. We shall call this mechanism, growth by ``substratum''.

\section{MEAN HEIGHT SPEED AND INTERFACE ACTIVITY DENSITY}
\label{sec:mh}

Replacing $L=1/\delta t$ and taking the limit $\delta t \to 0$,
Eq.~(\ref{ev}) becomes ${dh_i}/{dt}=G_i$.
Averaging over the lattice we obtain ($h=\langle h_i\rangle$)
\begin{equation}
\frac{dh}{dt}=
\langle 1-W_i \rangle + \langle F_i W_i\rangle\label{dhdt}\;.
\end{equation}
This equation allow us to identify the of two separate
contributions: diffusion $\langle 1-W_i\rangle$ and 
substratum $\langle F_i\,W_i\rangle$ \cite{note}. Yang and Hu
\cite{yang} defined two kinds of growth events: an event in which
the growth occurs at the chosen site (type $A$--defined by us
as substratum growth) and the event in which the growth occurs at
the adjoint site (type $B$--our growth by diffusion). They
counted, in numerical simulation, the events number $N_A(t)$ of type
$A$ and $N_B(t)$ of type $B$, in a time interval $L$. They did not
identify this terms as contributions to the mean height speed
(MHS). Notice that $N_A(t)\propto\langle F_i\,W_i\rangle$ and
$N_B(t)\propto\langle 1-W_i\rangle$ (see Figure~\ref{MGs}). We
shall see in Section~\ref{sec:ru} that the separation of those two
terms allows us to show how the diffusion enhances the roughness
near the critical value. The separation into two contributions for
all the quantities studied in this paper has never been done
before.

The substratum contribution can be expressed as
$f-\langle F_i\,(1-W_i)\rangle$, where
$f=\langle F_i\rangle$ is the IAD. We found an amazing numerical result:
\begin{equation}
\langle F_i\,(1-W_i)\rangle=p\;\langle 1-W_i\rangle\label{amazing}\;.
\end{equation}
We could not analytically obtain this result. Notice that $F_i$ and
$1-W_i$ are not independent, and that $f\neq p$ for $t>0$, as we shall see
bellow. Using the Eq.~(\ref{amazing}), the IAD is
\begin{equation}
f=p\,\langle 1-W_i \rangle + \langle F_i\,W_i\rangle\label{ff}\;.
\end{equation}
Figure \ref{IAD} shows both sides of this equation as function of time
showing that Eq.~(\ref{amazing}) holds. Notice the similarity between
Eq.~(\ref{dhdt}) and Eq.~(\ref{ff}).
Figure \ref{MGs} shows the diffusion and the substratum contributions as a
function of the time for various values of $q$.
At the initial time $dh/dt=f=p$. In the early time regime the substratum
contribution dominates the diffusion one, because $1-W_i$ is very small.
The substratum contribution dominates the behavior of $f$ and $dh/dt$ in
the early regime. As growth continues, the probability that growth will occur
by diffusion becomes larger; the diffusion contribution increases and
the substratum one decreases. This can be explained heuristically:
inactive sites generate a difference of heights greater than two between
any site and his neighbor, enhancing the growth by diffusion.
As time goes on, long chains of pinned sites are generated, slowing down the
diffusion contribution and hence the substratum one.
For $q<q_c$ these contributions, which in turn dominate, saturate to
equilibrium in the asymptotic regime; while, for $q\ge q_c$, both
contributions go to zero because the system becomes pinned. At the critical
value both contributions gives rise to a power law in the IAD and the MHS.
Notice that only at the critical value does a power-law scaling holds for
the MHS (see Figure~\ref{DHM}), which contradicts \cite{tang}.
This was shown for the roughness by Braunstein and Buceta \cite{brauns1}.

\section{ROUGHNESS}
\label{sec:ru}

From the Eq.~(\ref{ev}), the temporal derivative of the square interface
width (DSIW) is:
\begin{equation}
\frac {d w^2}{dt} = 2\,
\langle(h_i-\langle h_i\rangle)\,G_i\rangle\;.\label{dwdt}
\end{equation}
Replacing $G_i$ from Eq.~(\ref{gi}), the DSIW can also be
expressed by means of substratum and diffusion additive
contributions. The diffusion contribution is
\begin{equation}
2\,[\,\langle(1-W_i)\,\min(h_{i-1},h_{i+1})\rangle-
\langle 1-W_i\rangle\,\langle h_i\rangle\,]\label{dsiw_d}\;,
\end{equation}
and the substratum contribution is
\begin{equation}
2\,[\,\langle h_i\,F_i\,W_i\rangle - \langle h_i\rangle\,\langle
F_i\,W_i\rangle\,]\label{dsiw_s}\;,
\end{equation}
where the relation $\Theta(x-x')+\Theta(x'-x)-\delta(x-x')=1$ has
been used to derive the diffusion contribution. In Figure \ref{DW}
we plot both contributions as a function of time for various values
of q. At short times, the diffusion process is unimportant because
$\Delta h$ is mostly less than two. As $t$ increases, the behaviour
of this contribution depends on $q$. Notice, from
Eq.~(\ref{dsiw_d}), that the diffusion contribution may be either
negative or positive. The negative contribution tends to smooth
out the surface. Figure \ref{DW} shows that this case dominates for
small $q$. The positive diffusion contribution enhances the
roughness. This last effect is very important at the critical
value. At this value, the substratum contribution is practically
constant, but the diffusion contribution is very strong, enhancing
the roughness. This last contribution has important duties on the
power-law behaviour. We think that it is amazing how the diffusion
plays a dominant role in roughening the surface. To our knowledge
the strong effect on the roughness, at the criticallity, of the
diffusion contribution has never been proven before.

Generally speaking, the substratum roughens the interface while the
diffusion flattens it for small $q$, but the diffusion also roughens
the interface when $q$ increases. The diffusion is enhanced by
substratum growth. The growth by diffusion may also increase the
probability of substratum growth. This crossing interaction
mechanism makes the growth by diffusion dominant near the
criticality.

\section{CONCLUSIONS}
\label{sec:con}

We wrote the ME for the evolution of the height in the TL model.
The ME allows us to separate the substratum and the diffusion
contributions and to explain the great interplay between them. We
found that both contributions to the MHS are related in simple
way. We found an amazing numerical result that allows to derive
the IAD in a simple way. The analytical prove of this numerical
result is still open. All the quantities studied shows, the strong
interplay of the diffusion and the substratum contribution in the
dynamics. The substratum growth enhances the diffusion; increasing
the growth by diffusion may increase the probability of
substratum growth, and vice versa. This crossing interaction
mechanism makes both dominant at the criticality. The diffusion
contribution of the DSIW shows different behaviour depending of
$q$. In the intermediate regime, when $q$ is small, this contribution
is negative, smoothing out the surface. It is astonishing that as $q$
increases the contribution became positive, roughing the surface.
Finally, we are sure that other DPD growth models would permit
separation in two contributions with the same features of the TL
model.

\acknowledgements

A. D\'{\i}az-S\'anchez acknowledges financial support from the
INTERCAMPUS E.AL.'96 Program to attend to UNMdP.

\newpage
\begin{figure}
\caption{ln-ln plot of $\langle F_i\,W_i\rangle$ ($\Box$) and $\langle
1- W_i\rangle$ ($\bigcirc$) versus $t$. The parameter $q$ is (A) 0.51 (B)
0.539 (C) 0.6.}
\label{MGs}
\end{figure}

\begin{figure}
\caption{ln-ln plot of $f/p$ versus $t$. The symbols show the right-side of
Eq.~(4) (in units of $p$) compute as is explained in
reference [5]. The full curve shows the left-hand side of the same
equation (in units of $p$) where $f=\langle F_i\rangle$.
The parameter $q$ is 0.51 ($\Box$), 0.539 ($\bigcirc$) and 0.6
($\bigtriangledown$). The critical case shows that the IAD goes as
$t^{-\eta}$ with $\eta\simeq 0.40$.}
\label{IAD}
\end{figure}

\begin{figure}
\caption {ln-ln plot of $p^{-1}\,dh/dt$ versus $t$.
The parameter $q$ is 0.51 ($\bigtriangleup$), 0.539 ($\bigcirc$)
and 0.6 ($\bigtriangledown$). All cases shows the same behavior
in the early time regime. The subcritical
case shows that the MHS asymptotically goes
to certain constant. The critical case shows that the mean height goes
as $t^{-\beta}$. The supercritical case shows that the mean height is
asymptotically constant.}
\label{DHM}
\end{figure}

\begin{figure}
\caption{DSIW (full curve), and its diffusion ($\bigcirc$)
and substratum ($\Box$) contributions versus $\ln t$; for $q$
equal to $0.3$ (A), $0.539$ (B) and $0.6$ (C).}
\label{DW}
\end{figure}


\begin{thebibliography}{10}
\bibitem{note} To compute the Eq~(\ref{dhdt}) or any equation derived
from the ME we froze the simulation at a time t. For this configuration
we compute for the $i$-th site all the contributions to the growth of this
site in the next time $t+\delta t$ without changing the configuration. Then,
we average over the lattice and over realizations. This technique has been
employed in other systems, see \cite{nbl}
\end{thebibliography}
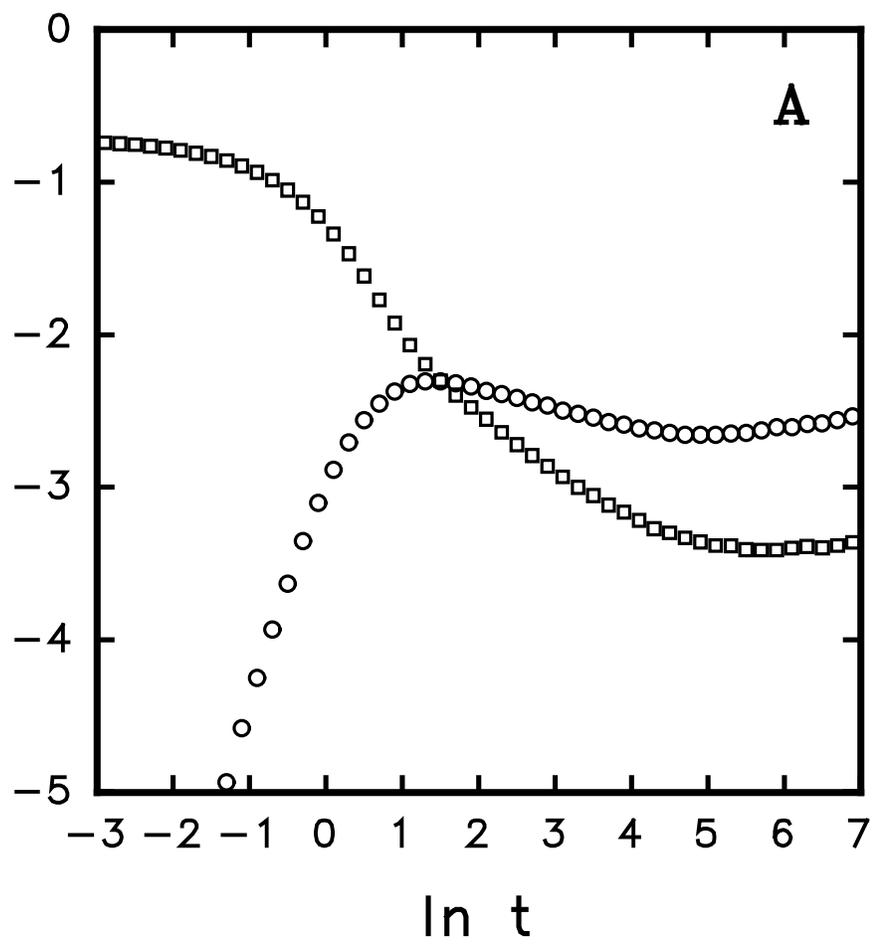
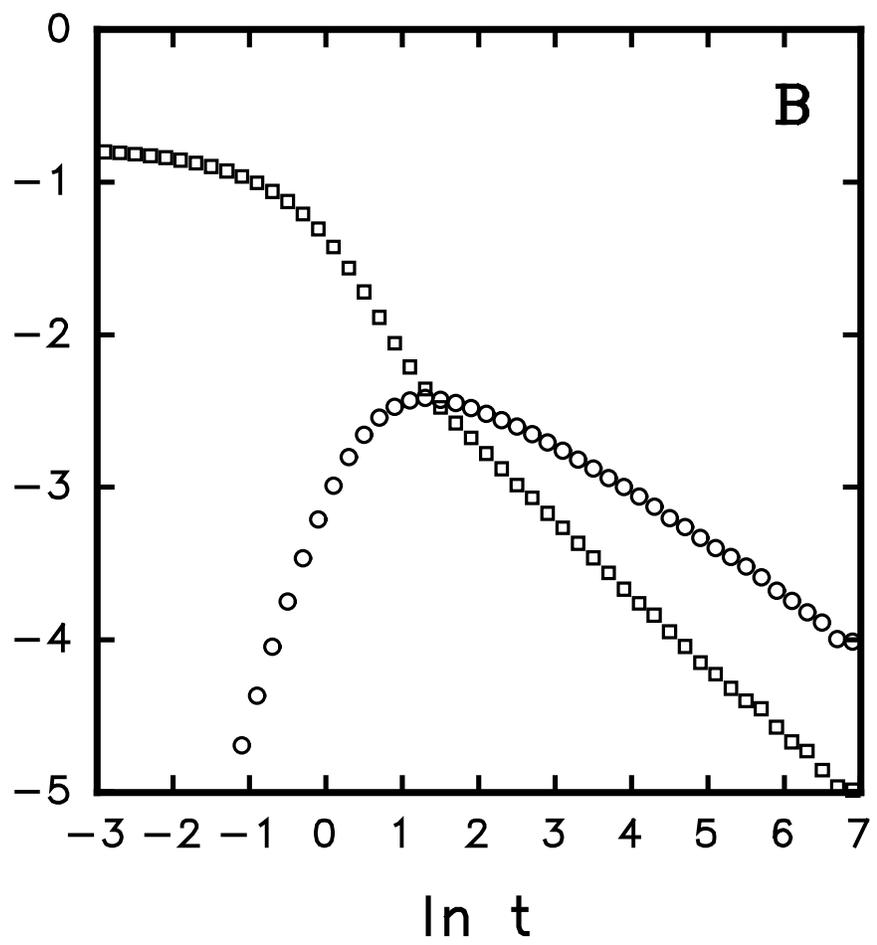
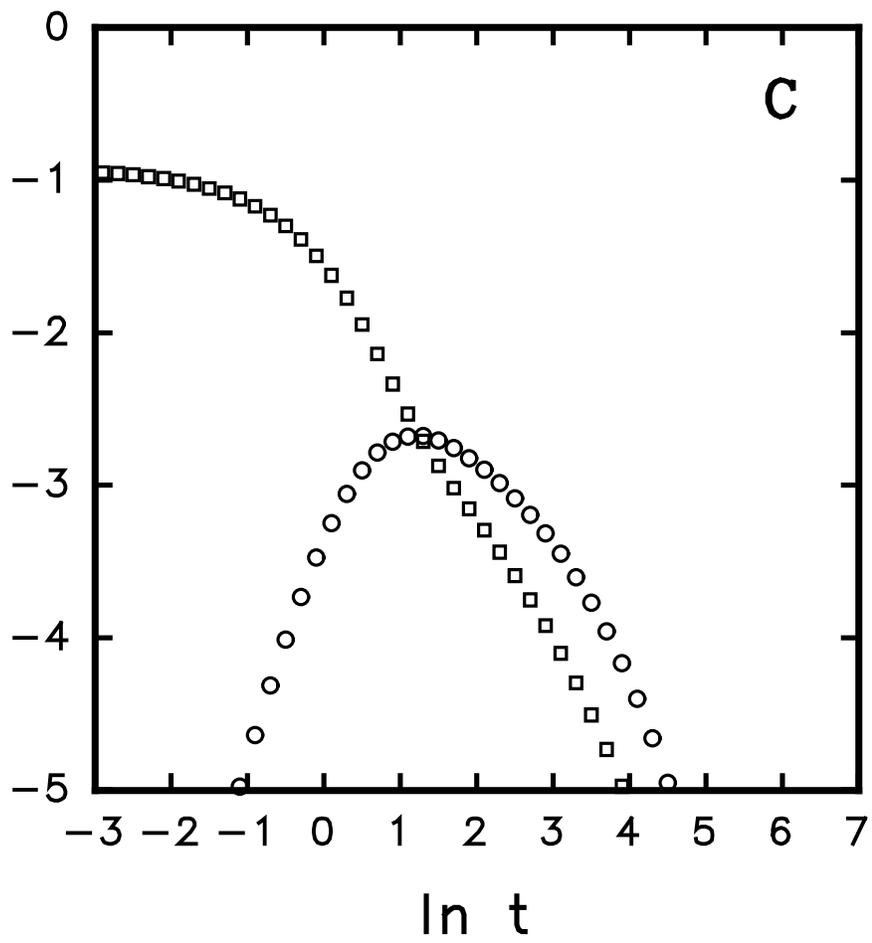
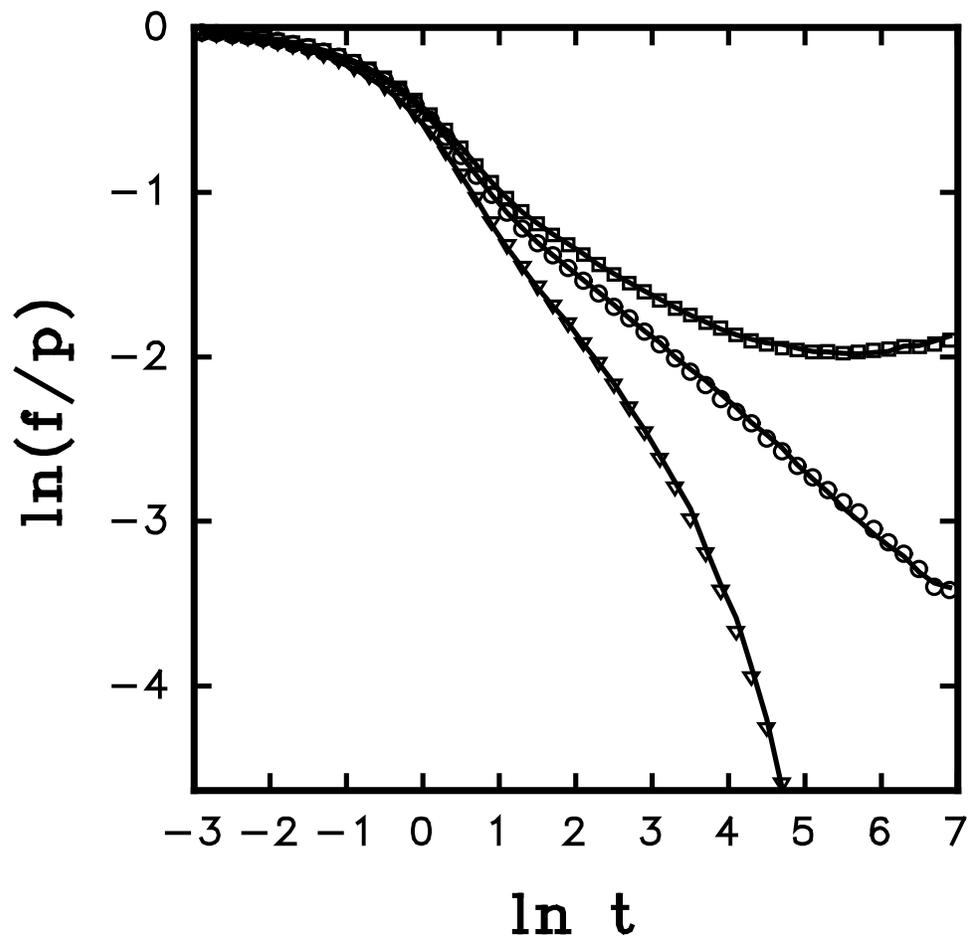
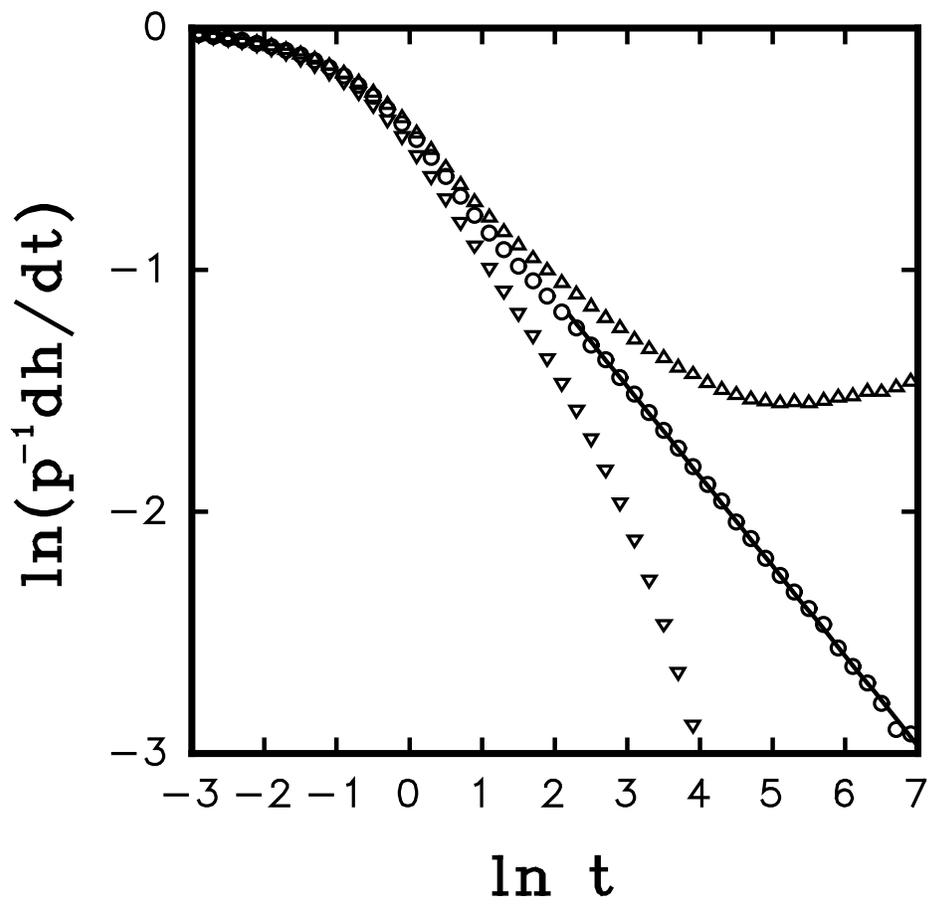
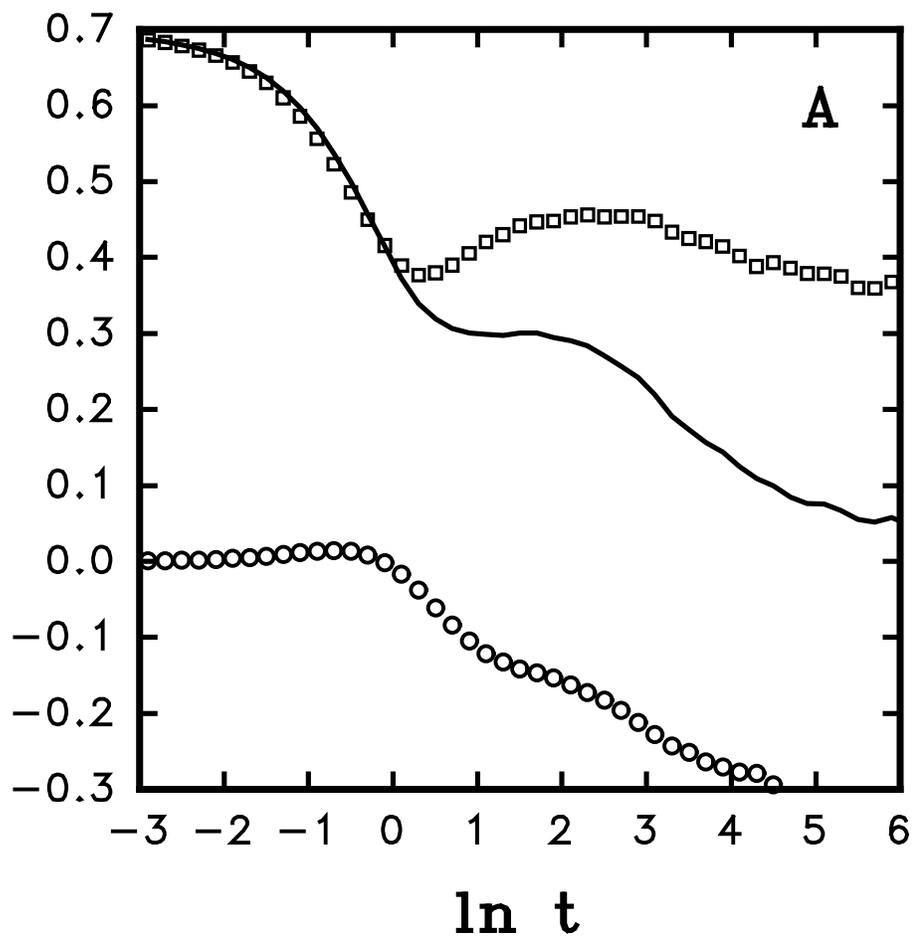
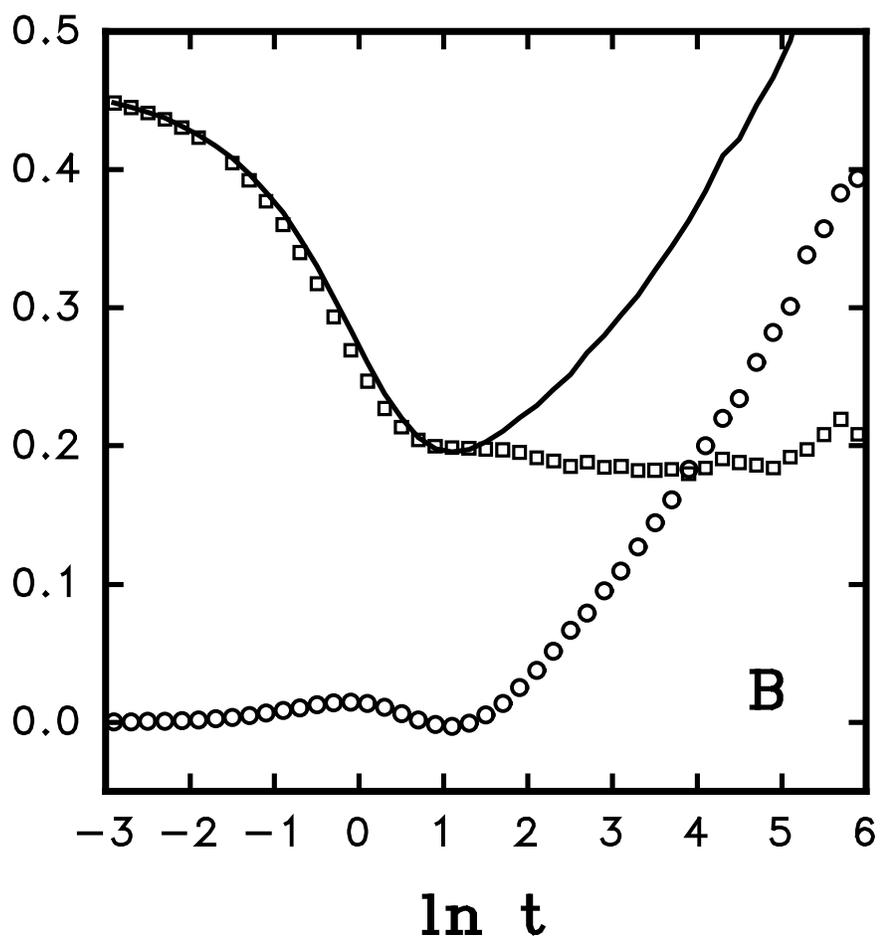
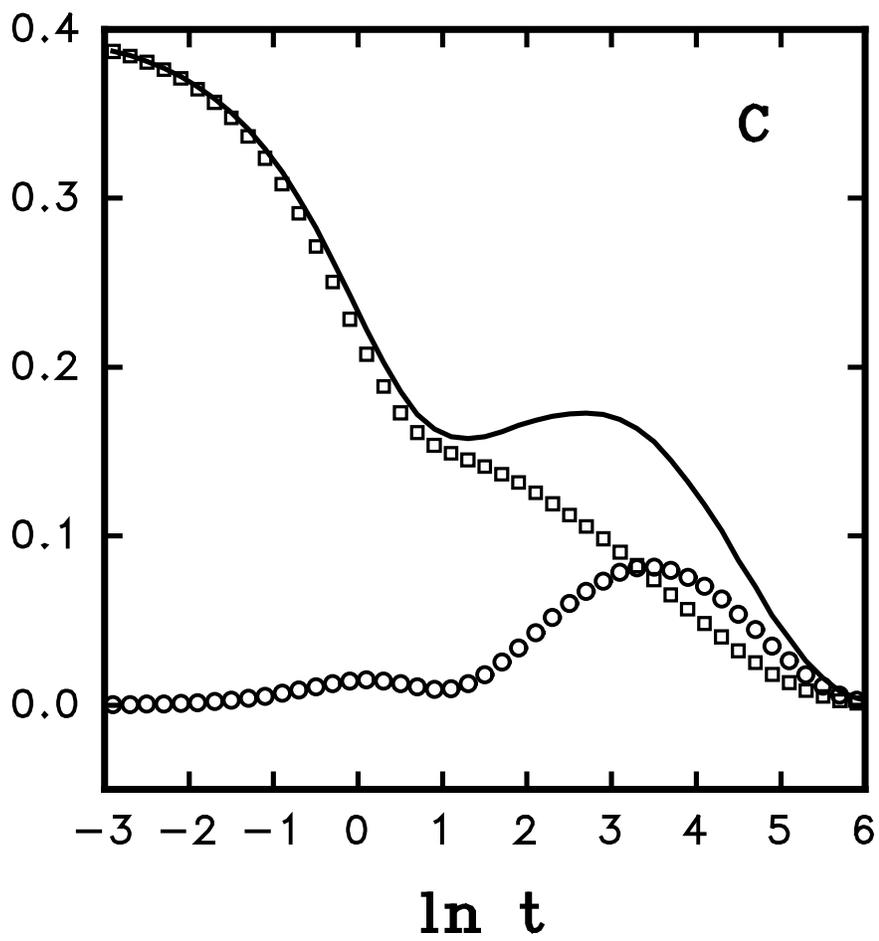
\end{document}